\documentclass{nature}

\usepackage{graphicx}
\makeatletter
\let\saved@includegraphics\includegraphics
\AtBeginDocument{\let\includegraphics\saved@includegraphics}
\renewenvironment*{figure}{\@float{figure}}{\end@float}
\makeatother

\usepackage[dvipsnames]{xcolor}
\usepackage{bm}
\usepackage{amsmath,amsfonts}
\usepackage[skip=2pt,font=footnotesize]{caption}

\title{Slow manifolds in recurrent networks encode working memory efficiently and robustly}

\author{Elham Ghazizadeh$^{1^*}$ \& ShiNung Ching$^{1,2,3}$}

\begin{document}

\maketitle

\begin{affiliations}
 \item {\footnotesize Electrical and Systems Engineering, Washington University in St. Louis, 1 Brookings Dr, St. Louis, MO 63130, USA }
 \item {\footnotesize  Biomedical Engineering, Washington University in St. Louis, 1 Brookings Dr, St. Louis, MO 63130, USA}
 \item {\footnotesize  Division of Biology and Biomedical Sciences, Washington University in St. Louis, 1 Brookings Dr, St. Louis, MO 63130, USA} 
\end{affiliations}

\begin{abstract}
Working memory is a cognitive function involving the storage and manipulation of latent information over brief intervals of time, thus making it crucial for context-dependent computation. Here, we use a top-down modeling approach to examine network-level mechanisms of working memory, an enigmatic issue and central topic of study in neuroscience and machine intelligence. We train thousands of recurrent neural networks on a working memory task and then perform dynamical systems analysis on the ensuing optimized networks, wherein we find that four distinct dynamical mechanisms can emerge. In particular, we show the prevalence of a mechanism in which memories are encoded along slow stable manifolds in the network state space, leading to a phasic neuronal activation profile during memory periods. In contrast to mechanisms in which memories are directly encoded at stable attractors, these networks naturally forget stimuli over time. Despite this seeming functional disadvantage, they are more efficient in terms of how they leverage their attractor landscape and paradoxically, are considerably more robust to noise. Our results provide new dynamical hypotheses regarding how working memory function is encoded in both natural and artificial neural networks.

\end{abstract}


Working memory (WM) is a temporary store that allows for active manipulation of information in the absence of external stimuli \cite{cowan2008differences}. Critical cognitive functions such as reasoning, planning and problem solving rely on working memory and thus its mechanistic basis is a key question in science and machine intelligence. Presumably, memory retention relies on an invariant latent neural representation of past stimuli \cite{chaudhuri2016computational}, but the precise nature of these representations and the dynamical mechanisms by which they are created in neural circuits remain enigmatic. 
Experimental and theoretical characterizations of working memory typically center on a delay period that occurs after stimulus presentation and before onset of a behavioral response or action \cite{romo1999neuronal,funahashi1989mnemonic, harvey2012choice, miller2018working}. Characterizations of neural activity during delay periods dichotomize into two broad categories: (i) persistent, tonic activity and (ii) time varying, phasic activity. In the former, neurons are tuned to relevant features of a stimulus and produce elevated and relatively constant activity throughout the delay \cite{riley2016role, wang2001synaptic, compte2000synaptic}. In the latter, neuronal activity fluctuates, ramping up and down during delay periods \cite{harvey2012choice, zhang2015dynamic,  murray2017stable}. Tonic and phasic paradigms have been observed in working memory tasks in animals \cite{zhang2015dynamic, murray2017stable} and artificial networks \cite{barak2013fixed,chaisangmongkon2017computing}.  However, the mechanisms underlying these descriptions and the reasons why one may manifest over the other in certain circumstances is far from clear.

 Understanding the network mechanism of working memory often revolves around the role of self-sustaining attractors, including discrete fixed points\cite{zylberberg2017mechanisms}, which correspond to neuronal activity patterns that are maintained indefinitely in the absence of exogenous stimuli or perturbation.  Tonic delay activity is thought to coincide with such attractors \cite{barak2014working,zylberberg2017mechanisms, deco2003attention}, thus allowing for stable maintenance of memory representations for potentially arbitrary lengths of time.  

 On the other hand, in the phasic hypothesis memory representations do not coincide with self-sustaining attractors. Instead, high-dimensional neuronal activity fluctuations may project onto a lower-dimensional latent space upon which an invariant representation is held during delay intervals \cite{druckmann2012neuronal, spaak2017stable}. For example, if the activity of a neuron gradually drops, the activity of another neuron increases to compensate for that drop. Thus, during delay, neural activity may traverse  a low-dimensional manifold corresponding to this invariant representation \cite{chaisangmongkon2017computing, bondanelli2020coding}.  
 
Disambiguating the above mechanisms requires deriving an understanding of the generative processes that give rise to time-varying, task-evoked neural activity. Ideally, we would be able to analytically characterize these mechanisms in a dynamical systems framework that could reveal the details of the attractor landscape embedded within neuronal networks. 

However, ascertaining dynamical systems models of biological networks is not straightforward, especially at a level of scale commensurate with networks thought to be relevant to WM, such as prefrontal cortex \cite{park2019dynamically,bauer1976delayed}. Here, artificial recurrent neural networks (RNNs) can form an interesting and potentially useful surrogate from which to derive mechanistic hypotheses. Such networks 
can be optimized in a top-down fashion to engage high-level cognitive tasks that include WM requirements \cite{hoerzer2014emergence, enel2016reservoir, pascanu2011neurodynamical,maheswaranathan2019universality, song2016training}.

Then, the emergent dynamics of the synthesized model can be analyzed and used to make arguments for or against different mechanisms, based on the predictive validity of the model outputs relative to actual brain activity. 

In this spirit, recent works have tried to reconcile the aforementioned hypotheses regarding persistent vs. transient delay activity in the context of WM \cite{nachstedt2017working, enel2020stable, cavanagh2018reconciling, orhan2019diverse}. Orhan and colleagues \cite{orhan2019diverse} optimized RNNs to perform several short-term memory tasks and they observed that both tonic and phasic delay activity could arise, depending on specific task details and optimization/learning parameters. Similarly, Nachstedt and colleagues \cite{nachstedt2017working} showed that the existence of both mechanisms simultaneously can mediate reliable task performance in the face of uncertain stimulus timing.
However, it remains unclear what factors sway RNNs to manifest one mechanism over another and, related, whether they carry different  functional advantages.

With regards to the last point, the ability of optimized RNNs to predict actual brain activity may depend crucially on certain restrictions regarding the optimization method that is used, e.g., by encouraging solutions that manifest connection motifs that are more biologically realistic \cite{sussillo2015neural,chaisangmongkon2017computing}.  Thus, using RNNs to build potentially explanatory hypotheses regarding neural circuit mechanisms likely requires careful consideration of the numerical optimization strategy used,  including hyperparameters and initialization schemes, as well as prior constraints on network architecture \cite{maheswaranathan2019universality, yang2019task}. Expanding on this idea, in this work, we pursue the top-down RNN approach to study potential mechanisms underlying WM function, training thousands of networks to perform an analytically tractable sequential, memory-dependent pattern matching task. To train our network, we modify the First-Order Reduced and Controlled Error (FORCE) \cite{sussillo2009generating} method by using a temporally restricted error kernel to confine error regression to occur only during brief intervals within each trial. The proposed framework blends trial-based reinforcement learning with first-order regression, thus obviating the need for a continual external supervisory error-signal.

Our premise is that this revised optimization framework, by leaving long epochs unconstrained, may allow for a wider range of possible emergent dynamics.  Indeed, by optimizing RNNs across different hyperparameters and initialization schemes within this framework, we identify a diversity of network mechanisms, each achieving the desired function, but varying in their key dynamical properties. We find that networks can embed predominantly asymptotically stable fixed points, stable limit cycle attractors, or a combination thereof. Most interestingly, we show here that there are two distinct mechanisms by which stable fixed points can be used to serve memory encoding, one leading to tonic activation and the other leading to phasic activation. We show that the latter, while unable to sustain memories over arbitrary lengths of time (i.e., wherein the model `forgets') nonetheless constitutes a more efficient and robust mechanism by which memories can be encoded.

\section*{Results}

\subsection{Working memory can be encoded via distinct dynamical mechanisms associated with tonic and phasic neural activation.}

We enacted a trial-based WM task involving sequential pattern matching (SPM) that exhibits working memory requirements (Fig. \ref{fig1}a). In our design, high-dimensional stimuli are encoded as bivariate random processes, such that the network is required to temporally integrate each stimulus and then store a latent representation of said stimulus for later processing. We optimized RNNs to perform this task by using a modified FORCE method \cite{sussillo2009generating} that included a temporally restricted error kernel. Here, regression occurs at two phases during each trial: (i) during memory/delay periods, wherein we promote the formation of an invariant latent linear projection from neural units nominally associated with maintenance of a memory representation; and (ii) at the conclusion of each trial, wherein we promote a linearly decoded output response signal (Fig. \ref{fig1}a,b).  All other temporal epochs are unconstrained, thus obviating the need to generate an error signal continuously throughout trials, which may overly constrain the dynamics \cite{yang2020artificial} (see also Methods for additional details and Supplementary Fig. 1).

We found that optimized networks could produce both tonic and phasic activity patterns during delay periods, as exemplified for two different networks of 1000 neurons in Fig. \ref{fig1}c. In order to study the dynamical mechanisms underlying these overt patterns we first used a numerical criteria on neuronal activity at the end of the delay period, $T_d$.  Specifically, we arrested trials at $T_d$ and forward simulated the networks autonomously to ascertain whether the activity was sustained at a fixed point (see Methods).  We identified four distinct dynamical mechanisms that could mediate working memory. In the case of tonic activation, network activity would indeed remain persistent, i.e., $\textbf{x}(t)$, the state vector of neuronal activity, would remain near $\textbf{x}(T_d)$ with $\left\|\dot {\textbf{x}} \right\|\simeq 0$, indicative of a fixed point attractor (Fig. \ref{fig2A}a). We refer to this mechanism as direct fixed point encoding (DFP). In the case of phasic patterns, $\textbf{x}(t)$ in the forward simulation would deviate from $\textbf{x}(T_d)$.  In some cases, the network would always settle at a \textit{different} fixed point from the memory representation (Fig. \ref{fig2A}b, termed indirect fixed point encoding, IFP), independent of the stimulus or network initial condition.  In other cases the network would always asymptotically approach a stable limit cycle attractor (Fig. \ref{fig2A}c, limit cycle encoding, LC). In a fourth case (not depicted), the network could asymptotically approach either a disparate fixed point or a limit cycle, depending on the stimulus realization (termed mixed encoding, see Supplementary Fig. 2). In total, we optimized 1524 network models, of which 703 were identified of the direct fixed point (DFP) mechanism, 534 were of the indirect fixed point (IFP) mechanism, 182 were of the limit cycle (LC) mechanism, and 105 were of the mixed (Mix) mechanism. Given their dominance in the emergent solutions, our subsequent attention will be on understanding the workings of the DFP and IFP mechanisms, though we will later also untangle the factors that cause each mechanism to arise over the others.

\subsection{Indirect encoding efficiently uses the network attractor landscape.}

The above findings suggests that key invariant structures in the network attractor landscape -- stable fixed points and attractive limit cycles -- determine whether and how delay activity takes on a tonic or phasic characteristic. To delve further into these mechanisms, we attempted to analyze how networks in each of the four categories leverage their respective attractor landscapes during the task. 

We began by linearizing the dynamics at the origin and using mean-field results \cite{schuessler2020dynamics} to establish lower bounds on the number of fixed point attractors manifest in the network attractor landscape. Fig. \ref{f3}a,b show how our four mechanistic categories break down along three key properties of spectra of the ensuing Jacobian matrix, where distinctions are readily observed. Most notably, the landscapes associated with direct fixed point encoding involve a greater number of fixed point attractors relative to indirect encoding. In support of this point, Fig. \ref{f3}c illustrates representative low-dimensional projections of network activity in each of the four mechanisms with stable fixed points overlaid (here, we restrict attention to the positive quadrant, see also Discussion). In DFP encoding (Fig. \ref{f3}c), the sequential stimuli move the trajectory between different fixed points (associated with memory representations), culminating in an output that is itself associated with a different fixed point (i.e., here a total of four fixed points are used in the service of the task). In contrast, the landscape for IFP encoding (Fig. \ref{f3}c) involves a \textit{single} fixed point that does not encode memories, nor does it encode the nominal output (though, it is approached asymptotically if networks are forward simulated autonomously after trial cessation). Thus, IFP encoding is able to maintain invariant representations during the relevant memory periods without relying directly on the presence of multiple fixed point attractors (see also Supplementary Fig. 3.)


\subsection{Indirect encoding uses slow manifolds to sustain memory representations}

Following from the above, IFP encoding appears to use the geometry of the stable manifolds of the single fixed point to maintain memory representations. Fig. \ref{f4}a illustrates the spectrum of the linearized dynamics about the fixed point in the previous IFP example encoding model, where we see many eigenvalues  near the imaginary axis, indicating the presence of slow, stable manifolds along which activity flows in a relatively invariant fashion.  These manifolds provide the opportunity for a low-dimensional latent representation of memory to be maintained, despite phasic activity (along the manifold). Indeed, because we encourage linearly mapped latent representations via our optimization method (see Methods), we know these manifolds have a planar geometry in the firing rate  activity variables. In contrast, Fig. \ref{f4}b illustrates the spectra resulting from linearization about two memory fixed points in a DFP model. Here we note that eigenvalues are relatively offset from the imaginary axis, indicating rapid convergence to the fixed point. This conclusion is supported in Fig. \ref{f4}c, which shows the relative proportion of delay periods in which neurons are in the saturated (nonlinear) vs. linear range of the activation function, i.e. $\text{tanh(.)}$ , for each model we trained. The much larger proportion of saturated neurons in DFP encoding indicates that the mass of eigenvalues for these models is relatively contracted and offset from the imaginary axis (see Methods and equation \eqref{jac}) and thus associated with fast decay to the fixed points.

To further understand the circuit-level details mediating the DFP and IFP mechanisms, we characterized the connectivity between neurons. We first noted that DFP encoding leads to an overall much greater distribution of connectivity weights between neurons relative to IFP (Fig. \ref{f4}d).  Next, we sorted neurons according to their peak activation (as in Fig. \ref{fig1}c) and examined their average pre-synaptic activity throughout the course of trials.  We found that neurons in DFP encoding exhibited highly structured synaptic tuning to different stimuli and memory periods, in contrast to IFP encoding (Fig. \ref{f4}e). Finally, we examined the bidirectional synaptic weight between `adjacent' neurons (ones with temporally sequential maximal activation).  Here, DFP exhibits no systematic connectivity structure, while IFP shows that neurons with similar peak activation times are more tightly coupled (Fig. \ref{f4}f). This latter point suggests that traversal along the slow manifolds is mediated by an internal sequential, `daisy chain' type of structure embedded within the trained IFP encoding network.




 
\subsection{Stable manifold encoding is forgetful, but robust.} 

We sought to better understand the functional advantages of the different mechanism types. In this regard, we interrogated networks by extending delay periods beyond the nominal training requirements, a form of increased memory demand. 
The main question here is how increasing the memory demand in this way would affect activity and consequently degrade task performance. Fig. \ref{f5}a illustrates the comparison of  neural activity patterns for DFP and IFP encoding categories (with extended delay equal to five times the nominal delay interval). For DFP encoding, regardless of the length of the extended delay, the neural activity is unaffected since the network uses fixed points as the invariant structure to encode memory traces. Consequently, after the extended delay ends and the network receives the second stimulus, task computations can be executed correctly. However, for IFP encoding, during the extended delay interval, neural activity gradually drops away which results in loss of function due to deviation from the `correct' activity pattern upon receiving the second stimulus. 
Fig. \ref{f5}b summarizes the deviation from the nominally `correct' post-delay neural activity as a function of delay extension for our two FP mechanisms, as well as LC and Mix. As expected, for DFP encoding  this deviation is near zero. In contrast, for  IFP the networks can tolerate extended delay up to \% 100 of the nominal delay, after which point performance gradually drops, i.e., the correct representation is `forgotten'.

 
To assay other functional aspects of these mechanisms, we examined how performance of our networks would tolerate the presence of a distracting noise added to the actual stimulus.  Here, we found a counterintuitive functional advantage of `forgetting,' relative to the DFP mechanism.  We specifically added uncorrelated noise of differing variance to the first of the two sequential stimuli and examined deterioration from the nominal `correct' neural representation at trial conclusion.  For values of noise variance that are less than the stimulus variance (vertical dotted line Fig. \ref{f5}c) IFC (and indeed LC) encoding are highly robust to perturbations, and indeed variances in excess of an order of magnitude greater than the stimulus can be tolerated. In stark contrast, DFP encoding is highly fragile with respect to distracting noise, with rapid and near-complete breakdown of the correct neural representation after modest perturbation (Fig. \ref{f5}c).  To understand this mechanism we carefully studied the trajectories in low-dimensional space in the presence of distracting noise (Supplementary Fig. 4), from which we ascertained that the distracting noise was essentially placing the trajectory in an erroneous basin of attraction, i.e., causing an incorrect memory fixed point to be induced. This result runs counter to classical Hopfield-type associative memory theory \cite{hopfield1982neural}, which presumes that basins are useful to rejecting noise and uncertainty.  Our finding, in essence, indicates that the high reliance on many fixed points for stable memory representations in DFP encoding makes this mechanism more susceptible to the temporal integration of noise (see also Discussion).



\subsection{Initial network properties dictate the emergence of different solution dynamics.} 

Finally, we sought to understand the factors prior to optimization that bias the emergent dynamics towards one type of mechanism versus another. We considered three main network properties: (i) the strength of connectivity, $g$,  (ii) the variance of feedback, $\sigma_f$, and (iii) the sparsity of the initial connectivity matrix.  We varied these parameters over their possible ranges. Fig. \ref{MonteCarlo} illustrates the effect of different parameterizations: for small values of $g$ the trainability of networks is poor, but improves significantly as $g$ increases. In other words, large random initial connectivity facilitates training, consistent with known results \cite{schuessler2020dynamics, sussillo2009generating}. For $g < 1$ the untrained network has one stable fixed point at the origin and the emergent trained dynamics tend to be of DFC or IFC encoding (Fig. \ref{MonteCarlo}a).  Interestingly, the  variance of feedback weights, $\sigma_f$ has a notable effect on the emergent dynamics; for large values of $\sigma_f$ the networks tend to form DFC models and as $\sigma_f$ decreases only IFC and LC models arise (Fig. \ref{MonteCarlo}b). The sparsity of initial connectivity matrix has no significant effect on the trainability of networks nor the emergent dynamics (Fig. \ref{MonteCarlo}c).

\section*{Discussion}

\subsection{Learning a diversity of dynamics for working memory function}

In this work we used a top-down optimization-based approach to investigate potential dynamical mechanisms mediating WM function. By training/optimizing RNNs using a modification of the FORCE regression method, we found four qualitatively different types of network dynamics that can mediate function. At a mechanistic level, these solutions are differentiated on the basis of the number of asymptotically stable fixed points manifest in the network vector field and, crucially, how those fixed points are leveraged in the service of the task. We note especially two solution types, one reflecting neural memory representations that are highly persistent corresponding to direct encoding at fixed points (i.e., DFP), versus the other where neural representations are transient and correspond to traversal along slow manifolds in the network state space (i.e., IFP). At the level of neural activity, DFP produces tonic sustained activity during delay periods, while IFP produces phasic, transient activity.


Our results are related to prior work that has shown that persistent versus transient encoding of memories can manifest in neural networks trained on different WM tasks and under different optimization/learning schemes \cite{orhan2019diverse}. Here, we choose to focus on a single, structured task in an effort to go beyond overt activity characterizations and carefully dissect the underlying dynamical mechanisms associated, namely the attractor landscape in the neural state space.
Doing so provides not only insight into potential generative circuit processes but also allows us to perform sensitivity analyses to ascertain nuanced functional advantages associated with the different mechanisms.

\subsection{Tradeoff between efficiency, memory persistence and robustness}
In particular, our results suggest an interesting balance between persistence and robustness of memory representations.  Specifically, the DFP mechanism resembles in many ways traditional associative memory attractor dynamics, in the sense of Hopfield networks \cite{hopfield1982neural}. Here, each memoranda is associated with a distinct, asymptotically stable fixed point.  On the one hand, such a mechanism is able to retain memories for arbitrary lengths of time.  Further, the dynamics with the attractor basins can nominally correct for small perturbations to neural trajectories at the onset of memory periods.  However, our results suggest that this latter classical interpretation breaks down when considering sequential, time-varying stimuli. In this case, perturbations to stimuli can accrue over time, causing neural representations to stray into errant basins of attraction, ultimately leading to failure of performance.

In contrast, in the IFP encoding mechanism, the network vector field exhibits a smaller number of fixed points that do not encode memoranda directly.  Rather, memory representations are formed from projection of neural activity along slow manifolds that are ostensibly shaped through optimization of the network vector field.  The fixed points here are, in essence, `shared' between memoranda. This mechanism turns out to be far more robust to time-varying stimulus perturbations.  There are likely two factors related to this robustness.  First, noisy perturbations may not be able to easily move trajectories off of the `correct' slow manifold.  Second, there are no competing attractors to absorb errant trajectories, as would be the case in the DFP mechanism.   
In total, the IFP encoding can be viewed as an overall more efficient use of neural dynamics wherein the lack of a persistent representation (i.e., `forgetfulness') is offset by both a lighter weight coding scheme in terms of the number of attractors deployed in the state space, leading -- perhaps paradoxically -- to more robust performance.  


\subsection{Shaping a landscape with few attractors} Expanding on the above point of efficiency, it is of note that the limit cycle and mixed mechanisms are most comparable to IFP in terms of the way in which they attractor landscape is used in the service of the task. In the LC mechanism in particular, the oscillatory cycle is not itself used to encode or sustain the memory, but rather shapes the landscape to create slow manifolds for encoding, similar to IFP. Thus, while the oscillation is not directly functional, it nonetheless is critical in establishing the `right' landscape for task completion. From an energetic standpoint, the indirect mechanisms are potentially less expensive since most neurons are inactive at any moment in time, in contrast to DFP encoding.

\subsection{Temporally restricted optimization promotes solutions that are compatible with observed dynamics  \textit{in vivo}}

Our results shed light on the different means by which recurrent networks can embed memory functions within their dynamics. Such a question is highly relevant to understanding how key machine learning and artificial intelligence constructs such as RNNs encode complex context-dependent functions \cite{maheswaranathan2019reverse}.  However, they also suggest mechanistic interpretations for actual WM circuits in the brain, given the seeming prevalence of phasic activity patterns during delay intervals observed \textit{in vivo} \cite{park2019dynamically}.  Indeed, it has been observed that neurons in memory-relevant regions such as prefrontal cortex do not necessarily maintain persistent activity throughout long delay periods, but rather may `ramp' on and off at systematic time points \cite{park2019dynamically}, as is compatible with our IFC mechanism. Further, in the IFC mechanism, most neurons are lightly saturated (Fig. \ref{f4}c), meaning that most neurons are within a linear regime, as thought to occur in actual neural circuits \cite{rubin2015stabilized, yang2020artificial}.

Notably, the IFP dynamical mechanism only arises after using the proposed temporally restricted error kernel.  Indeed, we found that using the native FORCE method without such a kernel leads to poor trainability; and further those networks that do manage to be trained are highly fragile to the extended delay and noise perturbations we considered.  This fragility ostensibly arises due to the latent outputs being overly constrained in this situation. Indeed, the choice of how to constrain these outputs throughout the task is somewhat arbitrary in the first place.  Hence, the temporally restricted error kernel may be allowing for the emergence of more naturalistic dynamics in our RNNs.


\subsection{Potential for enhanced fast learning and generalization}
An important technical caveat is that we have set up our RNNs to produce activity in the positive quadrant. Hence, our analysis focuses on characterization of the attractor landscape in that region of the state space.  However, because we consider an odd activation function, we know analytically that the fixed points analyzed in our networks have `mirror' negative fixed points that are not directly used in the service of the task, which means that these dynamical features are essentially `wasted' by construction and network design. A speculative hypothesis is that these fixed points may allow the network to more quickly learn a related task with minimal synaptic modification, i.e., by leveraging the mirror dynamics that are already embedded in the network. Such a concept is related to the idea of meta-learning \cite{wang2018prefrontal} and may be an interesting line of future study.

\begin{figure}[!h]
\begin{center}
\centerline{\includegraphics[scale=0.8]{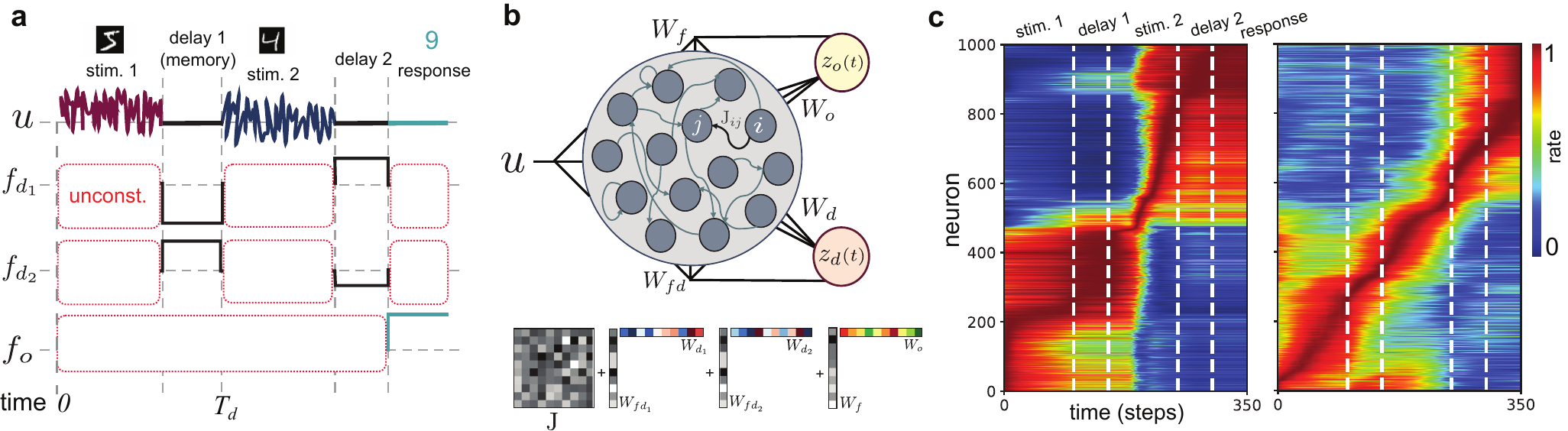}}
\caption{ \textbf{Optimizing RNN using modified FORCE to perform a sequential pattern matching (SPM) task.} \textbf{a,} A single trial of a SPM task, wherein low-dimensional random process representations of handwritten digit stimuli are followed by short delay intervals. The network is optimized to generate the correct `summation' output during a prescribed response interval. \textbf{b,} Schematic diagram of RNN architecture and low-rank structure added to initial connectivity $\bm{J}$. After the network receives input trials sequentially via input weights, it encodes the memory representation $z_d(t)$ and generates the task outputs $z_o(t)$. We use a rank 2 structure for encoding memory and a rank 1 structure for generating response. \textbf{c,} Tonic and phasic activity for two different networks. Activity patterns (normalized) of neurons are sorted by the time of their peak value.}
\label{fig1}
\end{center}
\end{figure}

\begin{figure}[!t]
\begin{center}
\centerline{\includegraphics[scale=0.8]{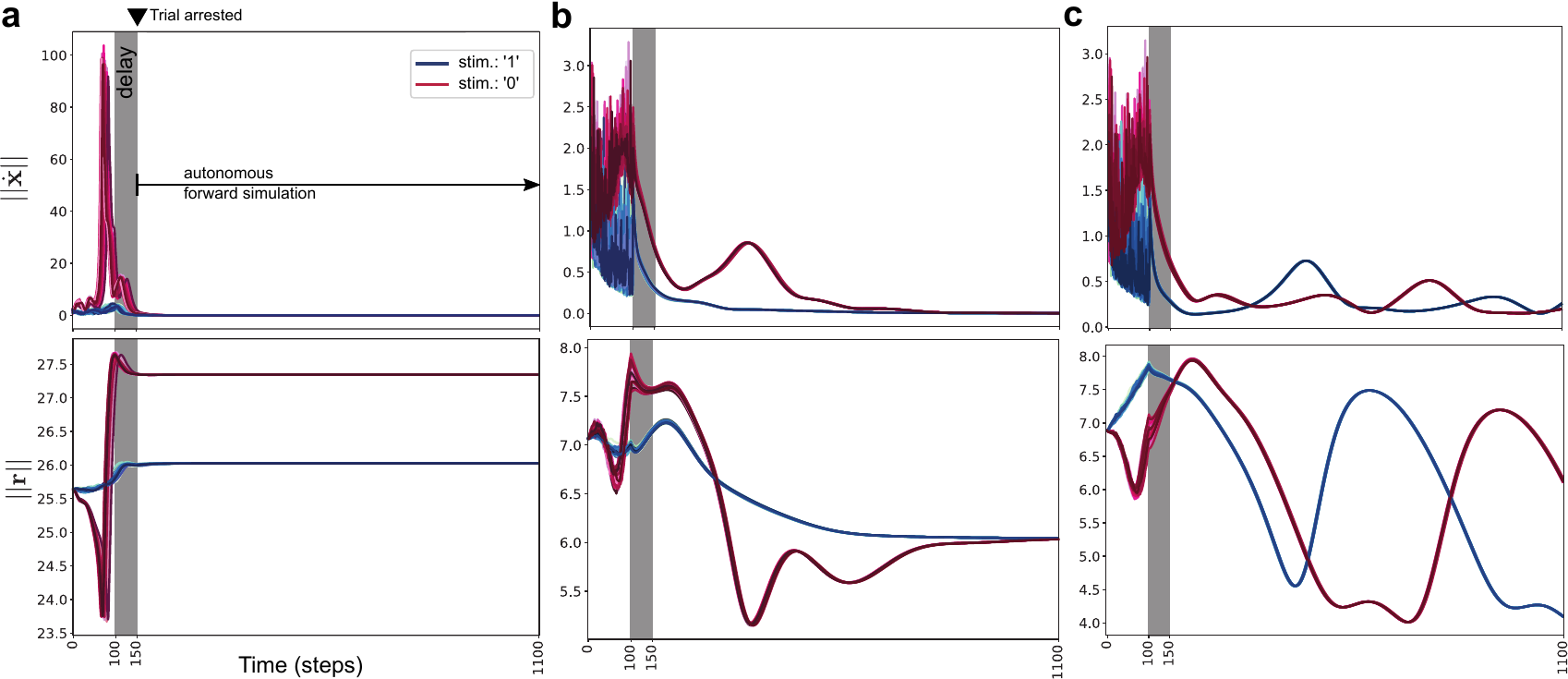}}
\caption{ \textbf{Forward simulation of network after delay to identify distinct dynamical mechanisms underlying WM }. \textbf{a,} Direct Fixed Point encoding (DFP), where the network uses fixed points to encode memory representations of each stimulus. \textbf{b,} Indirect Fixed Point encoding (IFP), where the network asymptotically settles at a fixed point but this fixed point does not correspond to a memory representation. \textbf{c,} Limit Cycle (LC), where the network asymptotically approached a stable limit cycle attractor.}
\label{fig2A}
\end{center}
\end{figure}

\begin{methods}

\subsection{Working memory task details.}

In this study, we considered a sequential pattern-matching task that takes into account key aspects of working memory tasks: stimulus processing, memory encoding and response execution \cite{park2019dynamically}. Our goal was to use a task of sufficiently low dimension as to be able to perform tractable and potentially illuminating post-hoc analysis on the emergent dynamics of RNNs.  In the proposed task, each trial consists of two random process stimuli that are sequentially presented and interleaved with delay intervals, followed by a brief response interval (Fig. \ref{fig1}). Each stimulus is a two-dimensional Gaussian process obtained in the latent space of a Variational Auto Encoder (VAE) trained on the MNIST dataset of hand-written digits. We designed the task pattern association rule to emulate summation, which differs from simple match or non-match tasks \cite{sakai2008task}. Specifically, to keep the dimensionality of the task low, we use two different stimuli resulting in 3 potential task outcomes (for summation). 

\subsection{Recurrent network model.}

We considered recurrent networks composed of $N$ nonlinear firing-rate units specified by:
\begin{equation}\label{init}
\tau \thinspace \dot{\textbf{x}}(t) = - \textbf{x}(t) + \bm{J}  \textbf{r}(t)
\end{equation}
where $\textbf{x} \in \mathbb{R}^N$ is the state vector and $\textbf{r }(t) =\text{tanh}(\textbf{x}(t))$ denotes the activity obtained via applying hyperbolic nonlinearity to the neuronal state variables. We set  the network time constant, $\tau=1$ for simplicity. Here, $\bm{J}$ is the (untrained) synaptic connectivity matrix with elements drawn randomly from a Gaussian distribution, i.e. $\bm{J}_{ij} \sim \mathcal{N} (0, \sigma_J^2)$. Specifically, we parameterize $\sigma_J^2 = \dfrac{g^2}{N}$, so that $g$ controls the strength of synaptic interactions. 
\subsection{Optimization method.}
Recurrent networks with fully random connectivity as in equation \eqref{init} have a rich dynamical repertoire and thus are capable of generating complex temporal patterns that are commensurate with spontaneous cortical activities \cite{barak2013fixed, sompolinsky1988chaos}. To make these networks learn the function of interest and thus perform the task, we first define two variables decoded from the network activity:
\begin{equation}
\begin{array}{l}\label{network0}
z_o(t) = \textbf{W}_o^T  \thinspace\textbf{r}(t)\\[5pt]
z_d(t) = \textbf{W}_d^T  \thinspace\textbf{r}(t)
\end{array}
\end{equation}
where $z_o(t)$ is a network output for generating responses, while $z_d(t)$ is a low-dimensional latent variable that is linearly decoded from neural firing rate activity, i.e. $z_d(t)= (z_{d_1}(t), z_{d_2}(t))$. In our network, invariant memory representations will be formed in this latent space. Optimization/learning proceeds by modifying the projection vectors $\textbf{W}_o \in \mathbb {R} ^{N \times 1}$ and $\textbf{W}_{d} \in \mathbb {R} ^{N \times 2}$. The network output $z_o(t)$ and dummy output $z_d(t)$ are fed back to the network via feedback weights i.e. $\textbf{W}_{f}\in \mathbb {R} ^{N \times 1}$ and $\textbf{W}_{fd} \in \mathbb {R} ^{N \times 2} $ , respectively. This results in modified synaptic connectivity:
\begin{equation}
\begin{array}{l}\label{network}
\dot{\textbf{x}}(t) =  - \textbf{x}(t) + (\bm{J} + \textbf{W}_f  \textbf{W}_o^T +  \textbf{W}_{fd}  \textbf{W}_d^T )  \thinspace \textbf{r}(t) + \textbf{W}_i \textbf{u}(t)
\end{array}
\end{equation}
The elements of $\textbf{W}_{f}$ and $\textbf{W}_{fd}$ are drawn independently from Gaussian distributions with zero mean and variance $ \sigma_f^2$. The network receives the exogenous input (i.e., stimulus) $\textbf{u}(t) \in \mathbb R^ {2 \times 1}$ via input weights $\textbf{W}_i \in \mathbb{R}^{N\times 2}$ (see Fig. \ref{fig1}b). This strategy effectively modifies the initial connectivity by addition of a low-rank component, allowing for more interpretable relations between the overall network connectivity and function \cite{mastrogiuseppe2018linking, schuessler2020dynamics}. Note that a minimal rank, i.e. rank 1, perturbation could be used, but it is known to induce high correlations between emergent fixed points, thus restricting the potential range of emergent dynamics \cite{beer2019one,schuessler2020dynamics}. Hence, to allow for a potentially wide range of solutions, we used a random connectivity plus rank 3 structure for the SPM task. 


 In our framework, optimization occurs only during the relevant temporal intervals in which these target signals are defined (Fig. \ref{fig1}a), which we term a temporally restricted error kernel.  When applying this kernel, the total error derived for a given trial is:
\begin{equation}\label{error}
 E(t) = \dfrac{1}{2}\int_{\mathcal T_{d}}  e_{d}(t)^2 dt + \dfrac{1}{2}\int_{\mathcal T_r}  e_{o}(t)^2 dt
 \end{equation}
where ${\mathcal T_{d}} $ and ${\mathcal T_{r}}$ are the temporal epochs associated with the two delay periods and response period, respectively (Fig. \ref{fig1}a).  Here,
\begin{equation}
e_{d}(t) = \left\| z_{d}(t) - f_{d}\right\|,
\end{equation}
where $z_d = (z_{d_1}, z_{d_2})$ and $f_d = (f_{d_1}, f_{d_2})$ and
\begin{equation}
e_{o}(t) = \left\| z_o(t) - f_{o}\right\|.
\label{fo}
\end{equation}
Here, $f_{d_1}$, $f_{d_2}$ and $f_{o}$ are scalar real numbers chosen prior to optimization to represent the 2-dimensional stimulus and the trial outcome. During the delay intervals in particular, a low error thus implies that the neural activity linearly maps to a constant, invariant representation (i.e., $z_{d} \in \mathbb{R}^{2\times 1}$). Activity during temporal epochs outside of the these periods do not impact the error. Optimization proceeds by modifying readout weights $\textbf{W}_o $ and $\textbf{W}_d$ to minimize these errors. 
 Within the temporal error kernel, we deploy the FORCE method for parametric regression in RNNs.  Here,  $\textbf{W}_o $ and $\textbf{W}_{d}$ are updated using recursive least squares  \cite{sussillo2009generating}. Briefly, to reduce $e_o(t)$, we obtain
\begin{equation}
\begin{array}{l}\label{force}
\textbf{W}_o(t) = \textbf{W}_o (t - \Delta t) - e_o(t) \bm{P}(t) \textbf{r}(t) \\[7pt]
\bm{P}(t) = \bm{P}(t - \Delta t) - \dfrac{\bm{P}(t-\Delta t) \textbf{r}(t) \textbf{r}^T(t) \bm{P}(t - \Delta t)}{ 1 + \textbf{r}^T(t) \bm{P}(t - \Delta t)\textbf{r}(t)}
\end{array}
\end{equation}
where $\bm{P}(t)$ denotes the approximate estimate for the inverse of the correlation matrix of network activities with a regularization term 
\begin{equation}\label{inv}
\bm{P}(t) = \int_{\mathcal T_r}   \textbf{r}(t) \textbf{r}^T(t) dt + \alpha \bm{I}_N
\end{equation}
where $\alpha$ is the regularization parameter and $ \bm{I}_N$ the identity matrix. In the same manner, to reduce  $e_d(t)$ we have

\begin{equation}
\begin{array}{l}\label{force}
\textbf{W}_d(t) = \textbf{W}_d (t - \Delta t) - e_d(t) \bm{P}(t) \textbf{r}(t). \\
\end{array}
\end{equation}
Note that we update the associated inverse correlation matrices during training intervals ${\mathcal T_{d}} $ and ${\mathcal T_{r}}$ (shown in Fig. \ref{fig1}a). In total, our training paradigm is a temporally regularized FORCE method that mitigates overfitting and in turn provides a potentially broader range of dynamical solutions to manifest. Indeed, it is known that optimizing RNNs using FORCE for a \textit{sequential} trial-based task (here, a pattern association task with memory requirement) prevents the emergence of multiple fixed points in optimized networks, and thus can overly constrain the range of possible solution dynamics \cite{beer2019one}.

\subsection{Dynamical systems analysis.}
The central theoretical question in our study pertains to analyzing the dynamics of our optimized networks. A first order question in this regard is to elucidate the landscape of attractors manifest in the network's vector field.  In equation \eqref{init}, the origin is always a fixed point associated with the Jacobian matrix  $\bm{J }$ (shifted by  $-\bm{I}$). 
To study the stability of the origin, we can thus consider the eigenvalues of the connectivity matrix.  It is well known that the eigenvalues of a random connectivity matrix $\bm{J }$ are distributed over a disk with radius $g$ for $N \rightarrow \infty$ \cite{sompolinsky1988chaos, rajan2006eigenvalue}. Thus, the stability of origin varies with these parameters. For $g < 1$ the origin is asymptotically stable, while for $g >1$ the origin is unstable, suggestive of potentially chaotic dynamics in the overall network. For the optimized networks with the rank 3 structure, the Jacobian matrix at the origin is $\bm{J}_T = \bm{J} + \textbf{W}_f  \textbf{W}_o^T +  \textbf{W}_{fd}  \textbf{W}_d^T$.   

Understanding the location and stability of fixed points away from the origin is harder to ascertain analytically.  Hence, we rely on a number of numerical procedures to identify these points.  To locate stable fixed points used for task computations, we arrest trials at relevant time moments, then forward simulate to ascertain the asymptotic behavior of the network. In one set of simulations, this forward simulation is carried out for trials arrested at the end of the first delay period. In a second set of simulations, it is carried out after trial conclusion. The forward simulation is carried out for ten times the nominal trial length, at which time we assume the network state is in a stationary regime, i.e., within an $\epsilon$ distance of either a stable fixed point or limit cycle.
We can perform additional linearization about stable fixed points that are discovered numerically in this way. Here, the eigenvalue spectrum of Jacobian matrix, $\bm{Q}$, at these non-zero fixed points, denoted $\textbf{x}^*$, is as follows
\begin{equation}
\bm{Q} = (\bm{J} + \textbf{W}_f  \textbf{W}_o^T +  \textbf{W}_{fd}  \textbf{W}_d^T) \thinspace \bm{R}^{\prime}
\label{jac}
\end{equation}
where $\bm{R}^{\prime}$ is a diagonal matrix with elements $\bm{R}^{\prime}_{ij} = \delta_{ij}  \textbf{r}^{\prime}_{i}$ with 
\begin{equation}
\textbf{r}^{\prime} = 1 - \text{tanh}^2 (\textbf{x}^*).
\end{equation}
Note that if the states are largely saturated at a fixed point (as in the case of DFP encoding, Fig. \ref{f4}c), then the entires of $\textbf{r}'$ are very small, which contracts the spectrum of $\bm{Q}$.

\subsection{Network connectivity analysis.}
To link network connectivity properties with the identified mechanisms, we first sorted neurons based on the time they reach their peak activity for each task trial. Then, we sorted elements of the optimized  connectivity matrix, i.e. $\bm{J}_T$ , using the same ordered sequence of neurons. We calculated the average pre-synaptic (or incoming) connections, i.e. $\bar {\bm{J}}_{i} = \dfrac{\sum_j^N {\bm{J}_T}_{ij}}{N}$, see Fig. \ref{f4}e. Moreover, we performed an analysis suggested by \cite{rajan2016recurrent}; we computed the average of diagonal and successive off-diagonal elements of sorted connectivity matrix, i.e. $|i-j| = c, c \in \{ 0, ..., N\}$, see Fig. \ref{f4}f, which indicates the strength of reciprocal coupling between neurons as a function of the temporal distance between their peak activation.

\subsection{Simulation parameters.}
In the task, we set the stimulus interval to 100 time steps, delay intervals to 50 time steps (except for the extended delay experiments) and response interval to 50 time steps. From the sequential bivariate random process stimuli, we trained a variational auto encoder on the MNIST digits 1 and 0.
We encoded the summation rule outcomes (i.e. 0, 1, 2) as 0.5, 1 and 1.5 (i.e., different values of $f_o$ in equation \eqref{fo}), respectively for training the networks. We encoded the latent outputs $f_{d_1}$ and $f_{d_2}$ as the two dimensional mean vector  for each digit representation, whenever that digit appeared prior to the delay period being optimized.

For all simulations the value of $\alpha$ is initialized to 1 and $\bm{P}$ was initialized to the identity matrix. 
The number of neurons was set as $N=1000$ and elements of $\textbf{W}_{o}$ and $\textbf{W}_{d}$ are initialized to zero. Input weights were drawn randomly from zero mean Gaussian distribution with variance 0.02. We set the time step, $dt$, for Euler integration to $0.1$. During training intervals, shown in Fig. \ref{fig1}a, we updated weights every 2 time steps. For four different initialization seeds we considered all possible combination of feasible values for $g$, $\sigma_f$ and sparsity (in Fig. \ref{MonteCarlo}). The value of $\sigma_f$ was chosen proportional to the size of network, i.e. $\sigma_f ^2\in \{ \frac{1}{1.5N}, \frac{1}{N}, \frac{1}{0.5N} , \frac{1}{0.1N}, \frac{1}{0.02N}, \frac{1}{0.005N},  \frac{1}{0.002N}, \frac{1}{0.001N}\}$. Training was terminated if the average root mean squared error between target and output was less that 0.01 for all trials. 

For exemplar networks used in figures \ref{f3}a,c and \ref{f4}a,b,d-f, for type DFP: $g=0.9, \sigma_f^2 = 1$ and sparsity is 0.2. For IFP: $g=0.9, \sigma_f^2 = 0.05$ and sparsity is 0.1. For LC: $g=0.9, \sigma_f^2 = 0.2$ and sparsity is 0.1. For Mix: $g=0.9, \sigma_f^2 = 0.1$ and sparsity is 0.1. The initialization seed is the same for these 4 exemplar networks. 

\subsection{Code availability.}
We performed simulations using Python and TensorFlow. All training and analysis are available on GitHub (\texttt{https://github.com/Lham71/Working-Memory-Modeling}). 

\end{methods}

\begin{addendum}
 \item This work was partially support by grants R01EB028154 and 1653589 from the National Institutes of Health and the National Science Foundation, respectively.  We gratefully acknowledge Dr. Larry Snyder and Charles Holmes for participating in discussions related to this work and for reviewing a preliminary version of this manuscript.
\end{addendum}

\begin{figure}[!t]
\begin{center}
\centerline{\includegraphics[scale=0.4]{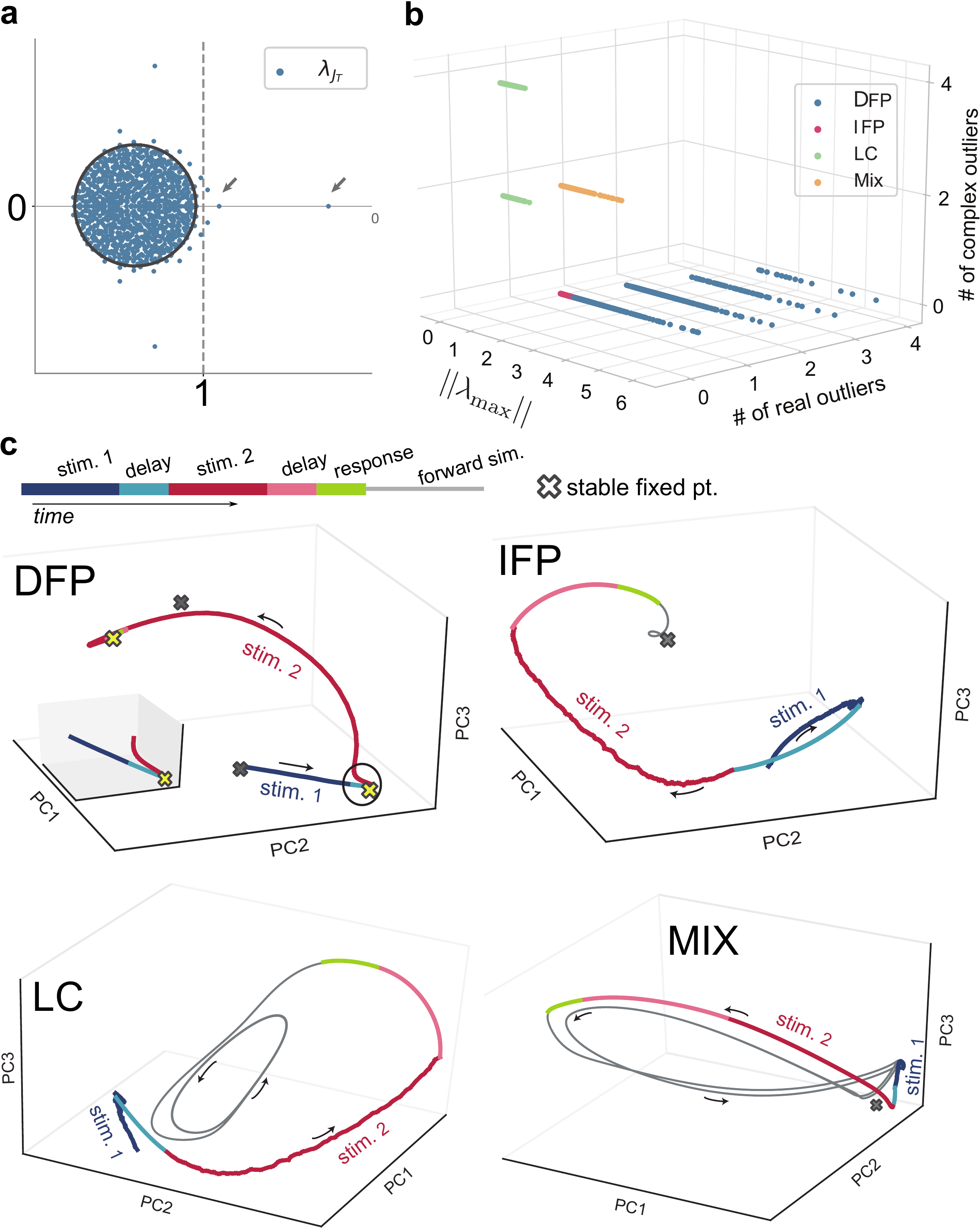}}
\caption{ \textbf{Attractor landscape for optimized networks} \textbf{a,} Eigenvalue spectrum (for an exemplar DFP network). The gray circle shows the radius of the theoretical eigenvalue spectrum of the initial connectivity matrix $\bm{J}$. After optimization, a set of outliers emerges in the eigenvalue spectrum. Here, the initial connectivity matrix is the Jacobian at the origin (shifted by $-\bm{I}$). \textbf{b,} Categorization of four distinct mechanisms along key properties of the network Jacobian evaluated at the origin. \textbf{c,} Attractor landscape and trajectory of exemplar task trials.  Three-dimensional neural trajectories are obtained via applying Principle Component Analysis (PCA) to 1000-dimensional neural activity from networks of each dynamical mechanism.  In DFP, the network creates 4 stable fixed points to solve the SPM task. For the displayed trajectory, the network uses two fixed points (shown in yellow) to directly encode the memory and trial output (the inset shows the area inside the circle). In IFP, the memory representation and trial output are encoded along the slow manifold of the single fixed point in the state space. In LC, the trajectories approach a stable limit cycle. For the mixed mechanism, both a stable fixed point and limit cycle are observed.}
\label{f3}
\end{center}
\end{figure} 
\begin{figure}[!t]
\begin{center}
\centerline{\includegraphics[width=\columnwidth]{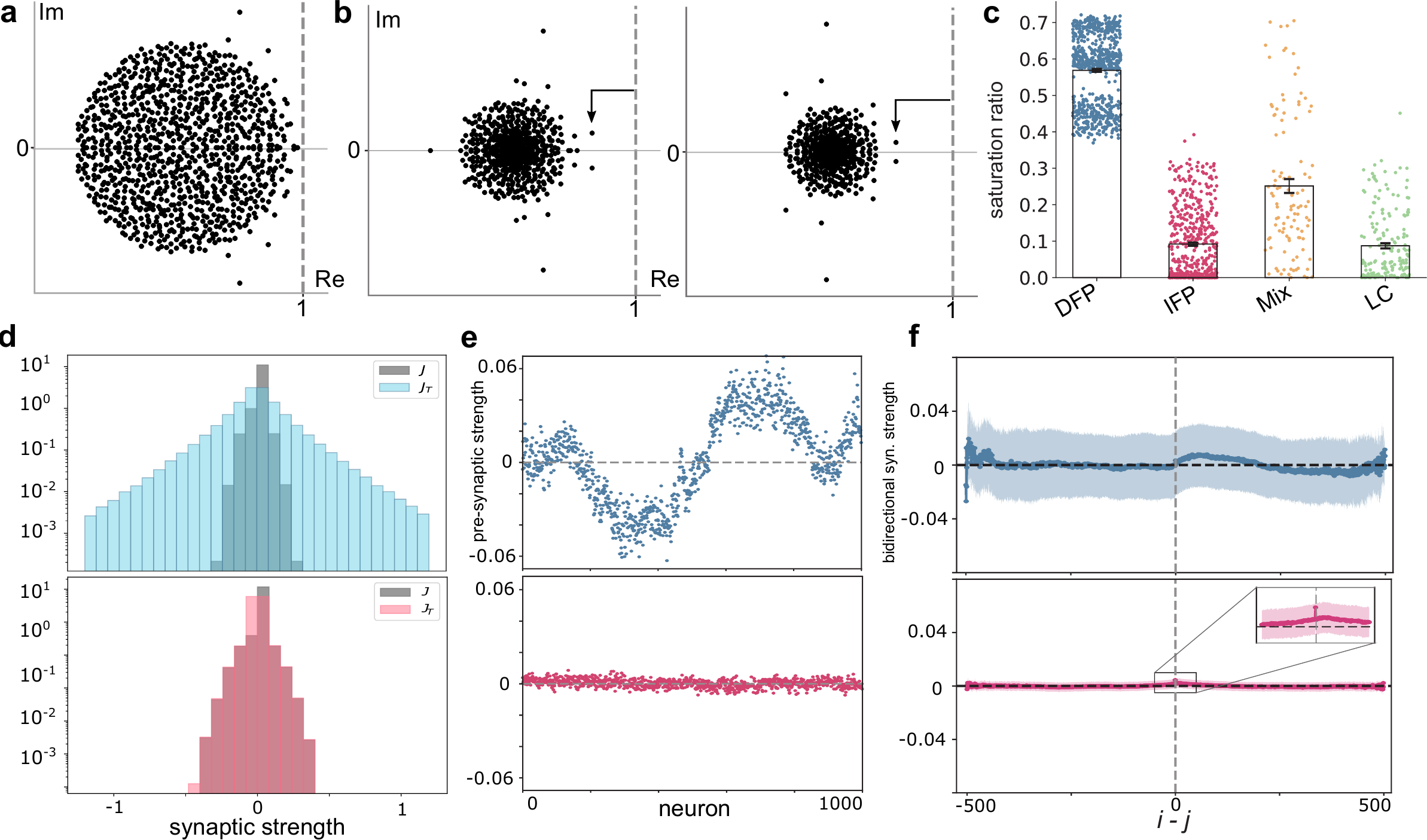}}
\caption{ \textbf{Eigenvalue spectrum at task fixed points and connectivity characterization.} \textbf{a,} Eigenvalue spectrum of Jacobian matrix at the single non-zero stable fixed point of IFP (shown in Fig. \ref{fig2A}b). \textbf{b,} Eigenvalue spectra of Jacobian matrix at memory fixed points of DFP (shown in Fig. \ref{fig2A}a). \textbf{c,} Saturation ratio (the ratio of neurons with activity in saturated range of activation function during memory interval (averaged over all trials)) for all networks simulated (across all four mechanisms). Standard error of the mean is depicted. \textbf{d,} Distribution of connectivity matrix entries (i.e., weights) before and after training for DFP (the top panel) and IFP (the bottom panel). \textbf{e,} Average pre-synaptic (incoming connections) strength sorted by peak activation of neurons (as in Fig. \ref{fig1}c) for DFP and IFP, respectively.  \textbf{f,} Comparison of mean and variance of elements of task connectivity matrix based on temporal distance of neurons. For IFP (the bottom panel) temporally adjacent neurons are more tightly coupled and a peak can be observed. (The inset shows this peak and $i,j$ denote neurons indices. ) }
\label{f4}
\end{center}
\end{figure} 

\begin{figure}[!t]
\begin{center}
\centerline{\includegraphics[width=\columnwidth]{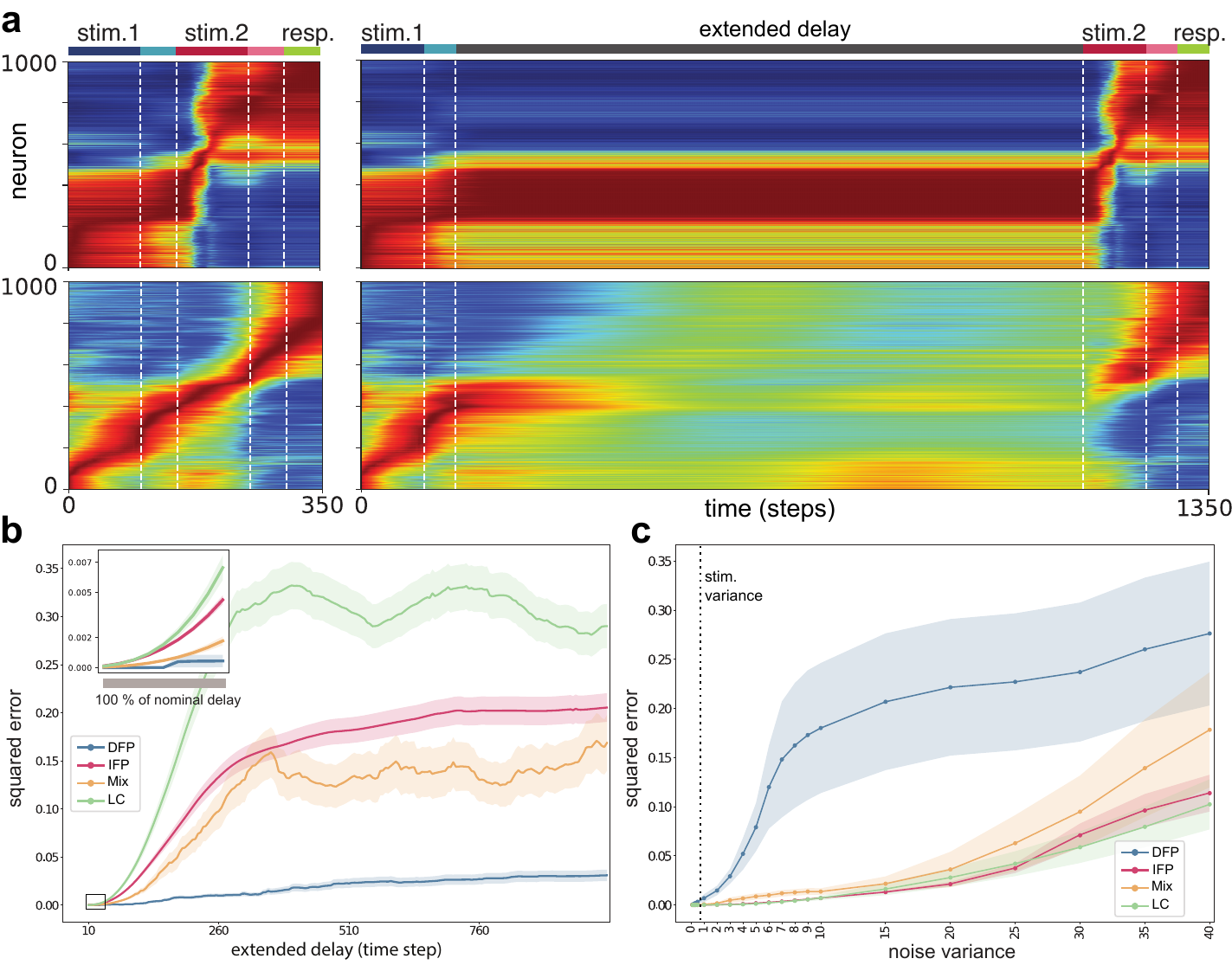}}
\caption{ \textbf{Functional advantages/disadvantages of each mechanism type.} \textbf{a,} Comparison of activity patterns before and after increasing memory demand for DFP (top panel) and IFP (bottom panel).  \textbf{b,} Summary of deviation from correct pattern of activity across different values of extended delay for all optimized networks. The squared error shows the difference between correct  and deviated trial outputs averaged over all trials and associated networks. \textbf{c,} Summary of deviation from correct pattern of activity across different values of noise variance for all optimized networks. \textbf{b} and \textbf{c} show that IFP, LC and mixed mechanisms are forgetful, but robust to sizable perturbations.}
\label{f5}
\end{center}
\end{figure} 


\begin{figure}[!t]
\begin{center}
\centerline{\includegraphics[width=\columnwidth]{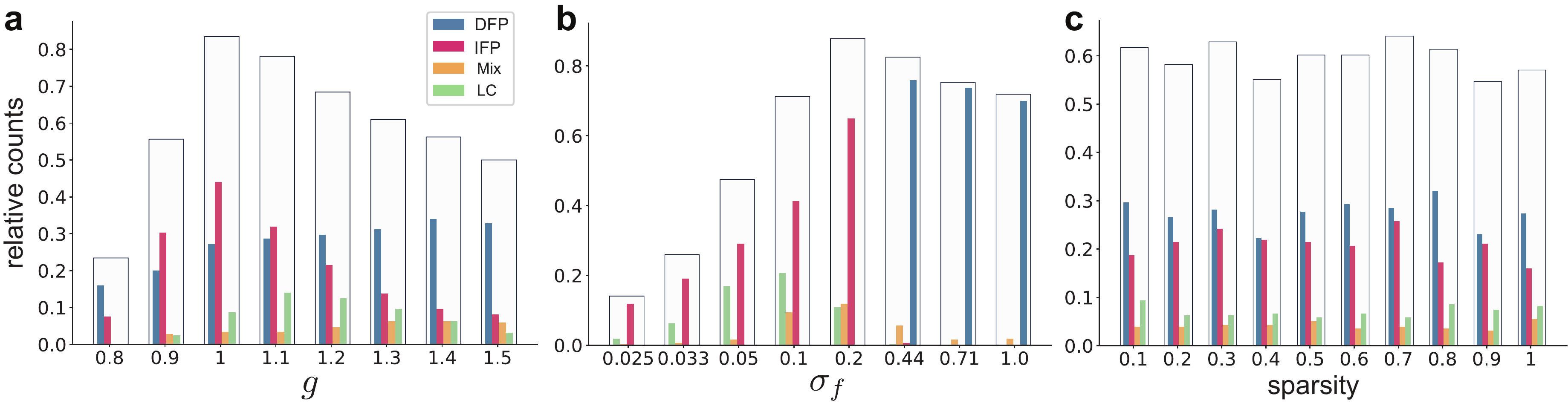}}
\caption{ \textbf{The effect of parameters prior to optimization on the diversity of the emergent solutions.} Relative count shows the number of trainable networks divided by the total number of networks for each specified value of parameters. Transparent bars show the relative count of trainable networks and the inner bars show the corresponding emergent types for each specified value of parameters.  \textbf{a,} The strength of connections within the initial connectivity matrix $\bm{J}$. \textbf{b,} The variance of feedback weights. \textbf{c,} The sparsity of $\bm{J}$. }
\label{MonteCarlo}
\end{center}
\end{figure} 


\newpage{\pagestyle{empty}\cleardoublepage}

\end{document}


\maketitle


\begin{figure}[!h]
\begin{center}
\centerline{\includegraphics[scale=0.8]{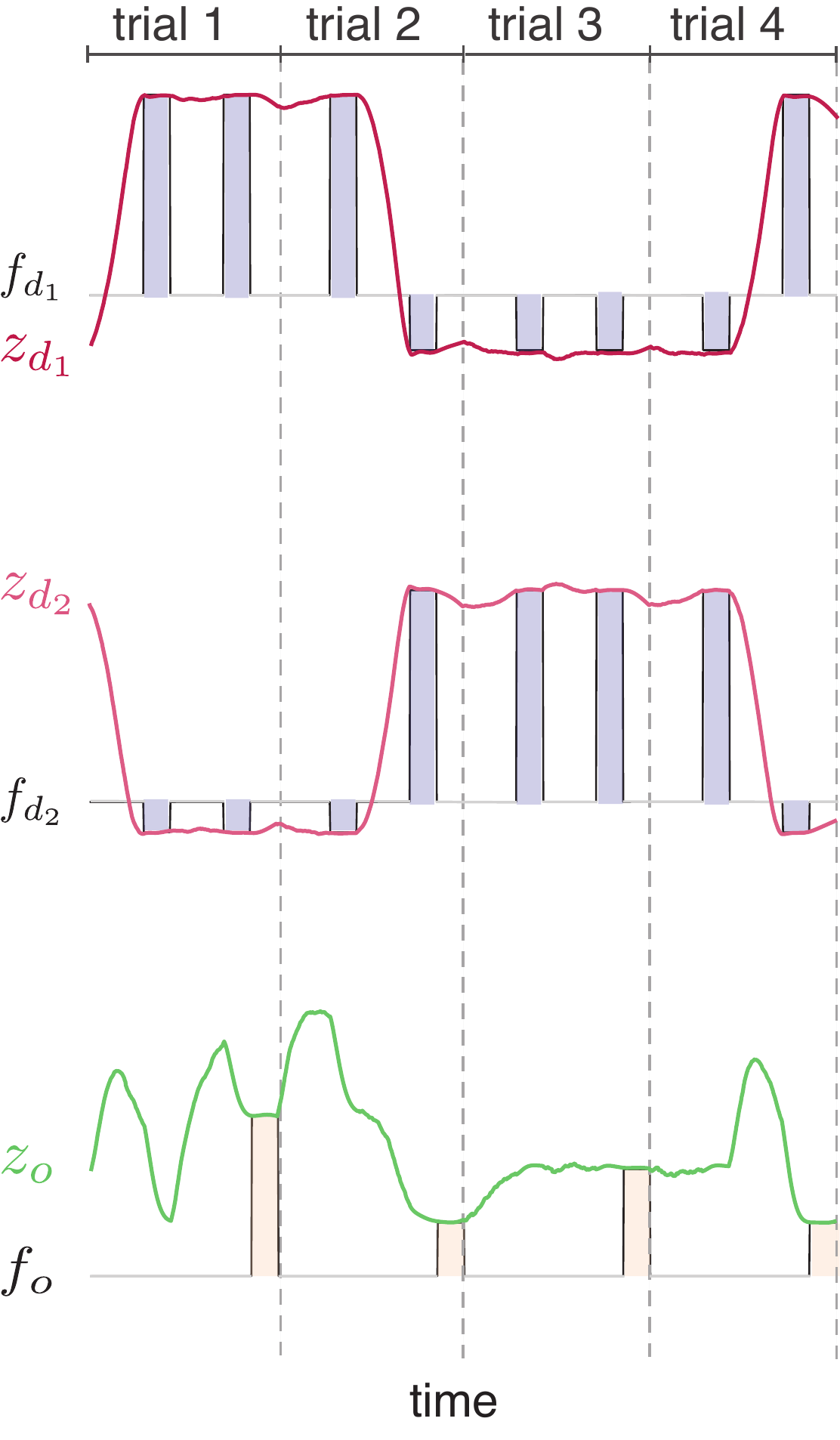}}
\caption{  \textbf{RNN outputs after optimization for performing SPM task}  Several concatenated trials of the task with outputs $z_d=(z_{d_1}, z_{d_2})$ and $z_o$ are shown. Using the modified FORCE method the network generates memory encodings during $\mathcal T_d$ and trial output during response intervals, $\mathcal T_r$ (shaded area).}
\label{s1}
\end{center}
\end{figure} 

\begin{figure}[!h]
\begin{center}
\centerline{\includegraphics[scale=0.6]{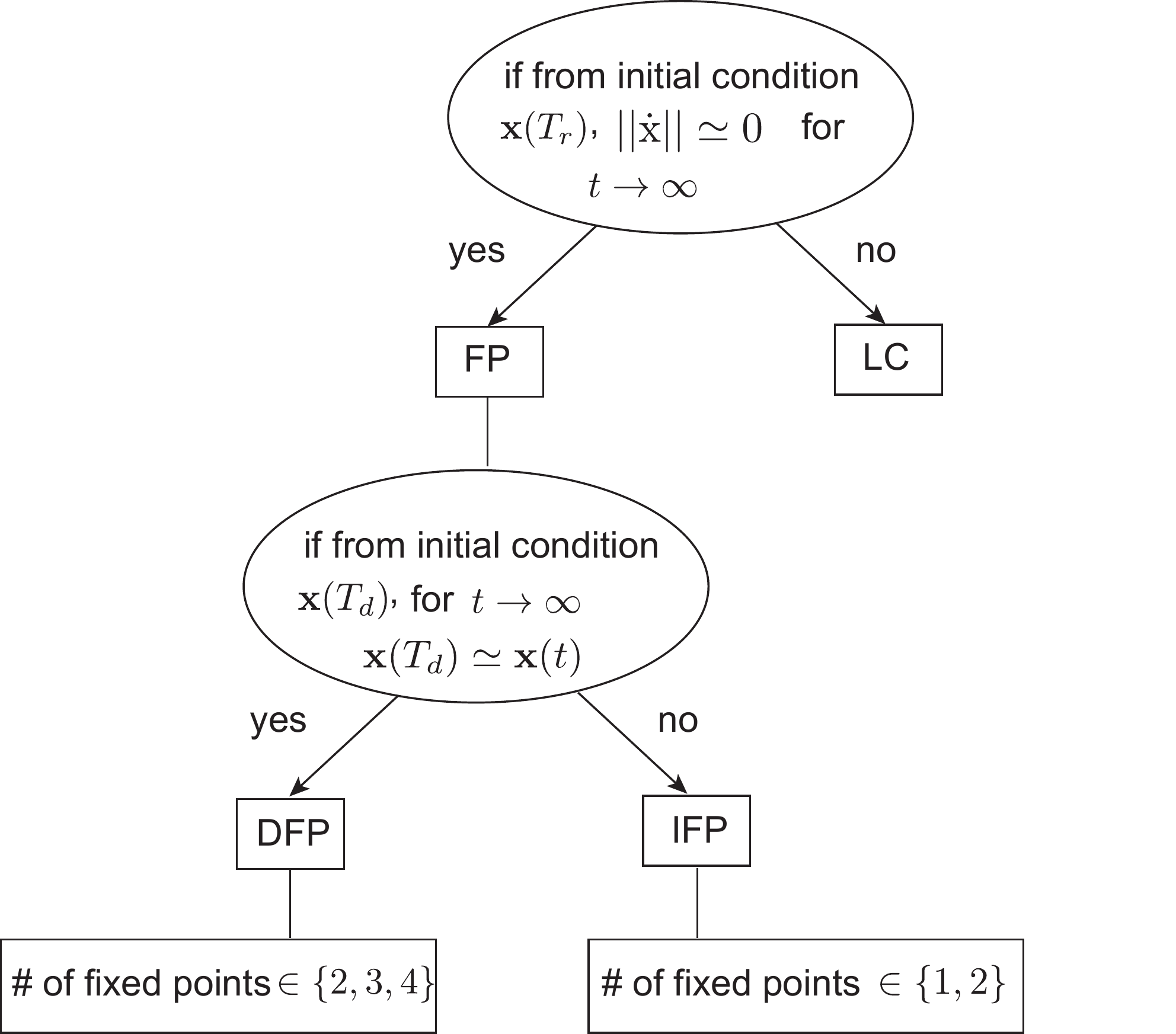}}
\caption{  \textbf{Flow chart} Categorization of dynamical mechanisms based on (i) the type of asymptotic attractor (either fixed point  or limit cycle) and (ii) whether delay periods correspond to a fixed point.}
\label{s2}
\end{center}
\end{figure} 

\begin{figure}[!t]
\begin{center}
\centerline{\includegraphics[scale=0.6]{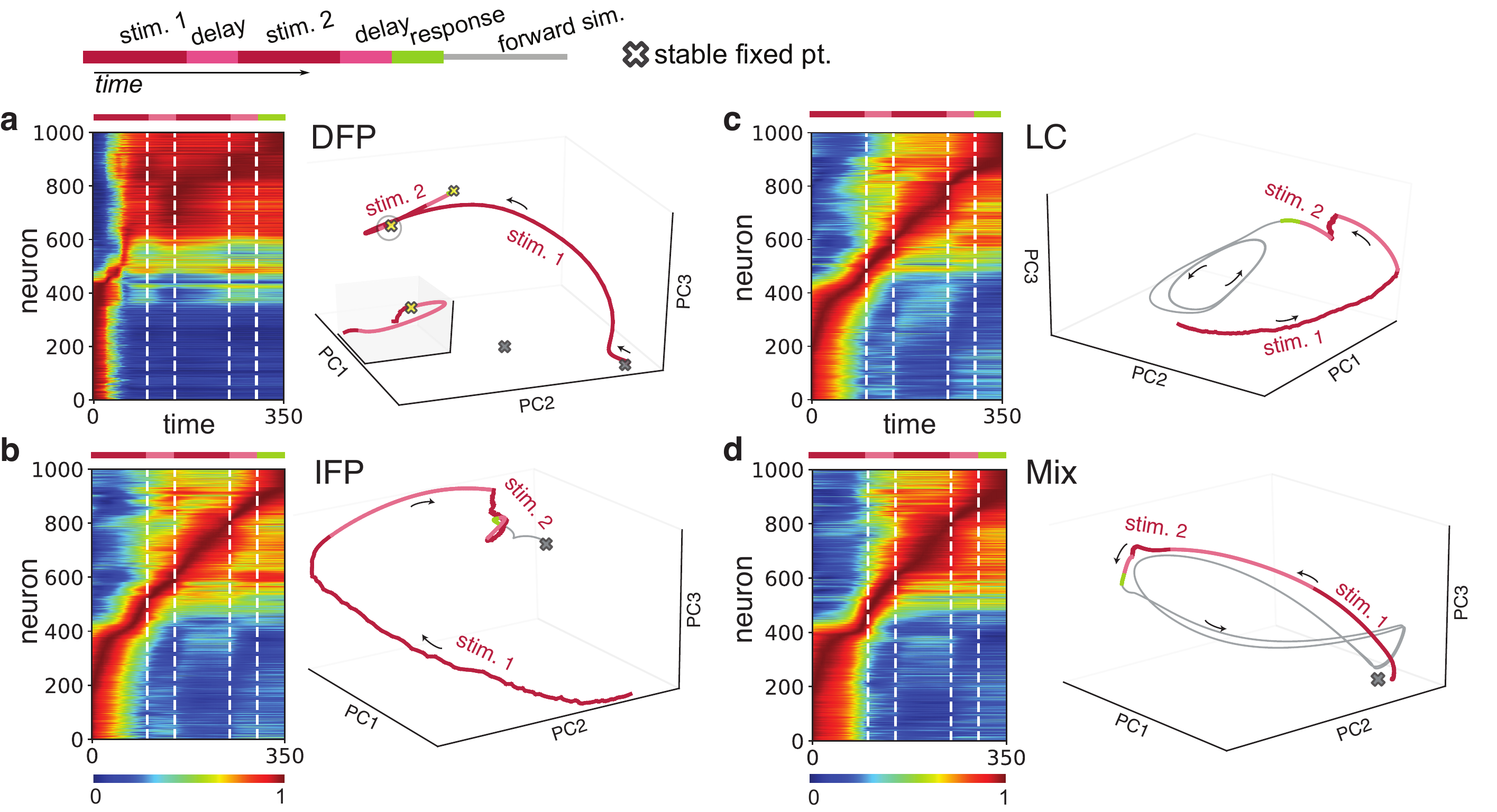}}
\caption{  \textbf{Neural activities and associated dynamical landscape for a single trial} For the same exemplar networks as in Fig. 3c, but a different set of stimuli (i.e., here, two realizations of the same digit are presented), neural activity and associated low-dimensional (PCA) trajectories are plotted. Note that PCA components are obtained for each exemplar network individually. The trajectories are color coded using the same scheme as the color bar on the top.  In DFP, the network creates four stable fixed points to solve the SPM task (the inset shows the area inside the circle). For the displayed trajectory, the network uses two fixed points (shown in yellow) to represent memory and trial output. In IFP, memory representation and trial output are encoded along the slow manifold of the single fixed point in the state space. In LC, the trajectories approach a stable limit cycle. For the mixed mechanism, both a stable fixed point and limit cycle can be seen. }
\label{fig2B}
\end{center}
\end{figure} 
\begin{figure}[!t]
\begin{center}
\centerline{\includegraphics[width=\columnwidth]{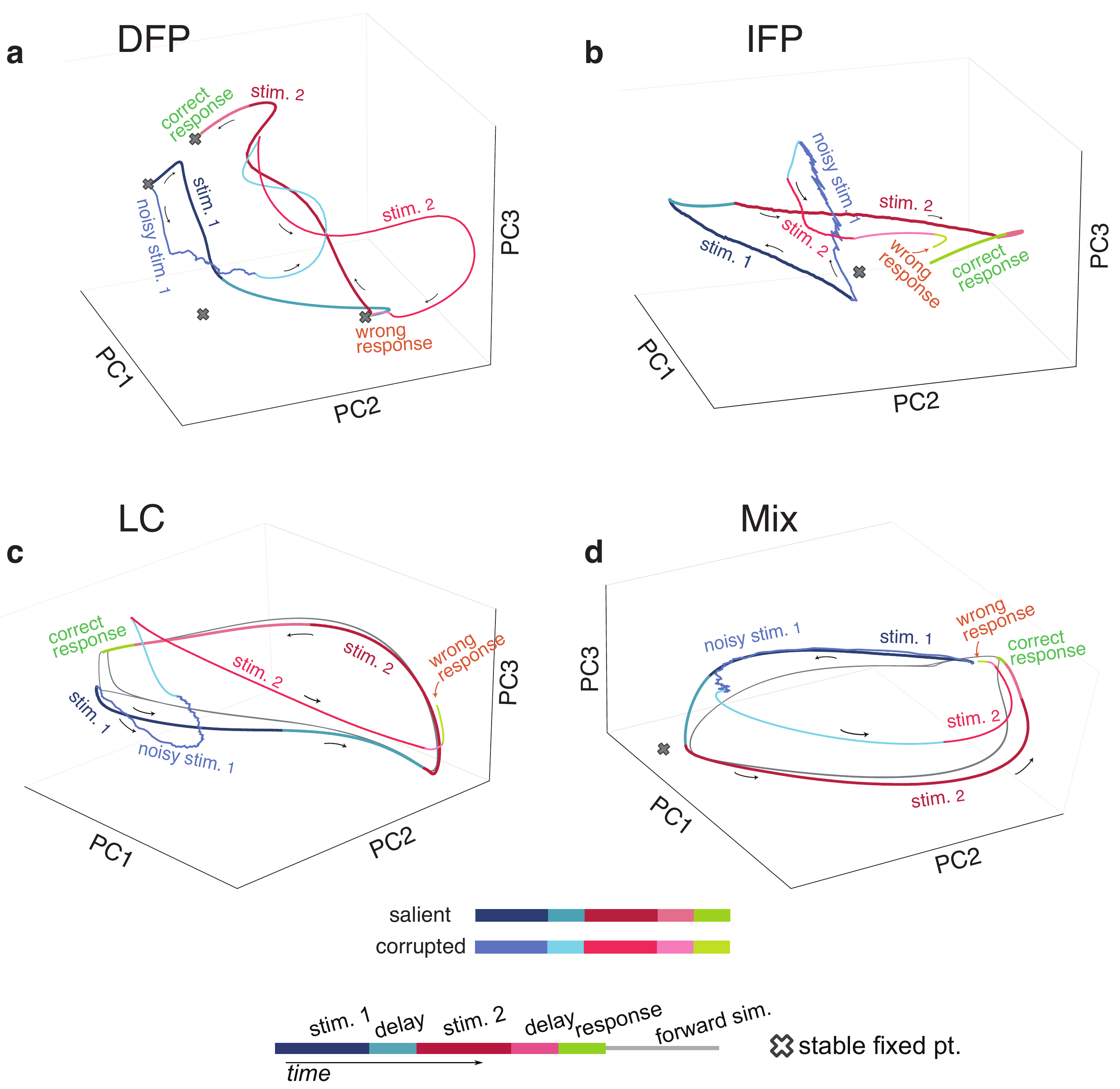}}
\caption{ \textbf{Neural trajectories of perturbed and salient trials} Plot show how noise corrupts the salient trajectory for all four mechanisms (same trial and initial condition). PCs are exclusive to each network. In DFP, distracting noise places the trajectory in an erroneous basin of attraction and thus the network generates an incorrect response; in IFP noise pushes the trajectory away from the `correct' slow manifold. }
\label{noise}
\end{center}
\end{figure} 
